\documentclass[letterpaper,twocolumn,prl,aps,superscriptaddress,amsmath,amssymb,floatfix]{revtex4-1}
\usepackage{mathptmx}
\usepackage[latin9]{inputenc}
\usepackage{color}
\usepackage{dcolumn}
\usepackage{amsmath}
\usepackage{amssymb}
\usepackage{graphicx}
\usepackage{esint}
\usepackage[unicode=true,
 bookmarks=true,bookmarksnumbered=false,bookmarksopen=false,
 breaklinks=false,pdfborder={0 0 1},backref=false,colorlinks=true]
 {hyperref}
\hypersetup{
 linkcolor=magenta,urlcolor=blue,citecolor=blue,pdfstartview={FitH}}

\makeatletter

\usepackage{textcomp}
\usepackage{epstopdf}

\usepackage{amsfonts}

\pdfpageheight\paperheight
\pdfpagewidth\paperwidth

\@ifundefined{textcolor}{}{%
 \definecolor{BLACK}{gray}{0}
 \definecolor{WHITE}{gray}{1}
 \definecolor{RED}{rgb}{1,0,0}
 \definecolor{GREEN}{rgb}{0,1,0}
 \definecolor{BLUE}{rgb}{0,0,1}
 \definecolor{CYAN}{cmyk}{1,0,0,0}
 \definecolor{MAGENTA}{cmyk}{0,1,0,0}
 \definecolor{YELLOW}{cmyk}{0,0,1,0}
}

\usepackage{xcolor}\usepackage{soul}
\definecolor{blue}{rgb}{0,0,1}
\definecolor{red}{rgb}{1,0,0}
\definecolor{green}{rgb}{0,1,0}

\makeatother

\begin{document}

\title{In-situ aligned all-polarization-maintaining Er-doped fiber laser mode-locked by a nonlinear amplifying loop mirror}
\author{Xiang Zhang}
\affiliation{Institute for Quantum Science and Technology, College of Science, National University of Defense Technology, Changsha 410073, China.}
\affiliation{Hunan Key Laboratory of Mechanism and Technology of Quantum Information, Changsha 410073, China.}

\author{Kangrui Chang}
\affiliation{Institute for Quantum Science and Technology, College of Science, National University of Defense Technology, Changsha 410073, China.}
\affiliation{Hunan Key Laboratory of Mechanism and Technology of Quantum Information, Changsha 410073, China.}

\author{Haobin Zheng}
\affiliation{Institute for Quantum Science and Technology, College of Science, National University of Defense Technology, Changsha 410073, China.}
\affiliation{Hunan Key Laboratory of Mechanism and Technology of Quantum Information, Changsha 410073, China.}

\author{Yongzhuang Zhou}
\email{y.zhou@nudt.edu.cn}
\affiliation{Institute for Quantum Science and Technology, College of Science, National University of Defense Technology, Changsha 410073, China.}
\affiliation{Hunan Key Laboratory of Mechanism and Technology of Quantum Information, Changsha 410073, China.}

\author{Yong Shen}
\affiliation{Institute for Quantum Science and Technology, College of Science, National University of Defense Technology, Changsha 410073, China.}
\affiliation{Hunan Key Laboratory of Mechanism and Technology of Quantum Information, Changsha 410073, China.}

\author{Hongxin Zou}
\email{hxzou@nudt.edu.cn}
\affiliation{Institute for Quantum Science and Technology, College of Science, National University of Defense Technology, Changsha 410073, China.}
\affiliation{Hunan Key Laboratory of Mechanism and Technology of Quantum Information, Changsha 410073, China.}

\date{\today}
\renewcommand{\figurename}{Fig.}

\begin{abstract}
Despite the wide applications for high-repetition-rate mode-locked fiber lasers, challenges persist in shortening the cavity length and coupling the  fiber collimators for most existing techniques. Here, we introduce a novel collimator alignment method and demonstrate an all-polarization-maintaining erbium-doped fiber laser that contains a nonlinear amplifying loop mirror with a repetition rate of 213~MHz. Compared to the conventional method, we achieve in-situ alignment of  the collimators in a simplified two-step process. Besides, through a comparison of the spectra from the output ports of the laser, we assess their quality and establish the spectral evolution relationships among these ports. It is found that, in addition to the widely believed large nonlinear effects, spectral interference also plays a significant role in spectral distortion. Moreover, a transition between different stability states is observed from the power variation of the single pulse.
\end{abstract}
\maketitle

\setcounter{figure}{0} 
\renewcommand{\thefigure}{\textbf{\arabic{figure}}}
\renewcommand{\figurename}{\textbf{Fig}}

\section{Introduction}   
Over the past twenty-five years, ultrashort pulse lasers have derived numerous applications, including precision ranging ~\cite{trocha2018,jang2023}, time and frequency transfer~\cite{gozzard2022}, materials processing~\cite{sugioka2021}, attosecond science~\cite{pupeza2021,midorikawa2022}, trace gas sensing~\cite{diddams2007,link2017}, and ultra-low noise microwave generation~\cite{nakamura2020,xie2017}. The generation of ultrashort pulses through passive mode-locking is typically achieved using intrinsic saturable absorbers such as SESAM~\cite{keller1990,alfieri2016}, carbon nanotubes~\cite{dai2020,huang2020}, graphene~\cite{wang2019,chen2014}, as well as artificial saturable absorbers like Kerr lens mode-locking (KLM)~\cite{spence1991,feng2021,zheng2021}, nonlinear polarization rotation (NPR)~\cite{inaba2006,wang2011,grelu2012}, nonlinear optical loop mirror (NOLM)~\cite{doran1988,szczepanek2014}, and nonlinear amplifying loop mirror (NALM)~\cite{kuse2016,hansel2017}. The NALM-based lasers offer simultaneously the advantages of all-polarization-maintaining (all-PM) fiber optics, easy mode-locking self-starting, and low cost. Therefore, they are among the most popular at present. However, in their high-repetition-rate version, shortening the cavity length and coupling the collimators still present challenges.

There have been significant advancements in optimizing the NALM-based fiber laser. In order to solve the problems of difficult mode-locking self-starting and low repetition rate in the Figure-8 lasers, the Figure-9 laser based on two polarization beam splitters (PBSs) was proposed~\cite{hansel2017}. This scheme allows for easier achievement of mode-locking self-starting in a shorter cavity length. However, problems arise when attempting to further shorten the cavity length. For example, reduced loop asymmetry prevents mode-locking and multiple fusion-splicing points in the fiber loop results in an unnecessarily long tail fiber. In order to address these issues, an integrated wavelength division multiplexer (WDM) collimator was invented~\cite{wang2011}. It is effective when using non-PM ytterbium-doped fiber but does not yield significant results when using erbium-doped fiber with lower gain~\cite{gao2018}. For the purpose of avoiding coupling the laser output with an additional collimator, a fiber coupler was added to the Figure-9 laser to achieve an all-in-fiber frequency comb for space applications~\cite{probster2021,lezius2016}. To simplify the coupling system, a single collimator structure was proposed to avoid coupling two orthogonal collimators that require complex six-axis alignment stage adjustments~\cite{fellinger2019}. In addition, the optical wedge pairs and the collimating reference surfaces are introduced to couple the collimators, making the system more compact and stable~\cite{yang2022,zhao2024}. Each of the above method offers unique advantages, however to our knowledge, hardly any reported method is able to combine them. What's more important, existing methods only allow for collimator coupling when the NALM loop is open. An in-situ and more efficient alignment method is desirable when coupling the collimators.

Each output port of the NALM-based fiber laser has its own specific characteristics. The experimentally observed spectra of the pulses from the two ports of the fiber loop show three-peak and single-peak structures, respectively, and these two pulses are also different in the time domain~\cite{zhang2023,jiang2016}. In the Figure-9 lasers, the centers of spectra at the reject port and output port are complementary~\cite{laszczych2022}. There are also differences in the background noise of the spectra of these ports~\cite{liu2019}. By studying the characteristics of these output ports from the same source, we can better understand their evolution relationships and select the best output port for the lasers. When the pump power of a passively mode-locked fiber laser exceeds a certain threshold, the single pulse operation becomes unstable and a multi-pulse or noise-like pulse regime can be realized. In the multi-pulse regime, the multistability states and the transitions between them can be observed by changing the pump power, which is termed hysteresis phenomena~\cite{li2022,komarov2005,tang2005}. In the single-pulse regime, these phenomena may occur, but to our knowledge, no experimental observations have been reported yet.

In this work, we propose an in-situ alignment method that combines some advantages of the above optimization methods, such as simplified coupling and potentially shorter cavity length. By integrating the alignment system into the laser structure itself, we demonstrate an all-PM mode-locked laser erbium-doped fiber based on a NALM. Besides, we measure and compare the output characteristics of the four output ports of the laser, including the typically neglected transmission output port of the PBS in the fiber loop of the Figure-9 laser, and provide the evolution relationship between them. Additionally, from the change in output power, a transition between two stability states is observed in the single-pulse regime.

\section{Experiment setup and Alignment method}
\begin{figure}[!t]
	\centering
	\includegraphics[width=0.3\textwidth]{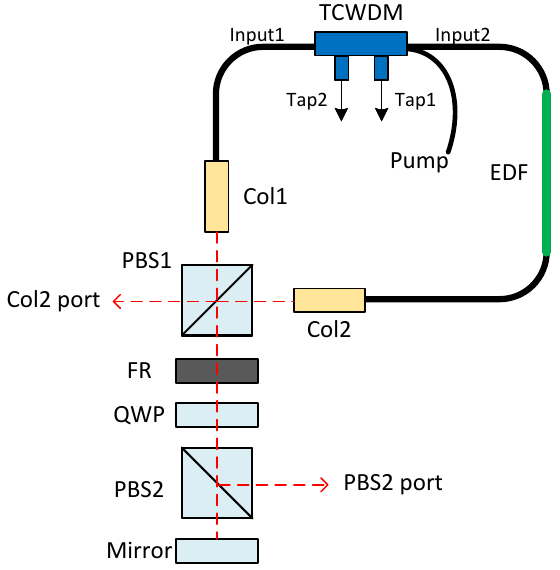}
	\caption{Experimental setup of the mode-locked fiber laser. Col1 and Col2, collimators; EDF, erbium-doped fiber; TCWDM, tap coupler hybrid with WDM; PBS1 and PBS2, polarization beam splitters; QWP, quarter-wave plate; FR, Faraday rotator. Four output ports: Tap1, Tap2, Col2 port and PBS2 port.}
	\label{fig1}
\end{figure}

The schematic of the all-PM erbium-doped fiber laser mode-locked by a NALM is shown in Fig.~\ref{fig1}. The working distances of the two collimators (Col1 and Col2) are 80~mm. The polarization of the slow-axis pulses output from Col1 and Col2 is parallel to the $p$ and $s$ states of the PBS1, respectively. These two orthogonally polarized light beams interfere on the PBS2 after passing through a non-reciprocal phase shifter composed of a Faraday rotator (FR) and a quarter-wave plate (QWP), showing a saturable absorption effect. The fiber loop consists of a 340~mm passive fiber (PM1550) and a 392~mm erbium-doped fiber (PM-ESF-7/125, Nufern). The group velocity dispersion (GVD) of these fibers is negative. We choose to use a low gain negative GVD erbium-doped fiber instead of a high gain positive GVD erbium-doped fiber (Er80-4/125-HD-PM, Liekki), for the following reasons: (1) The former can occupy more than half of the total fiber length, while the latter cannot because of the need to maintain a negative net cavity dispersion. (2) Solitons require less energy to sustain compared to breather solitons. The smaller difference in core diameter between negative GVD erbium-doped fiber and passive fiber results in lower fiber splicing loss.

\begin{figure}[!t]
	\centering
	\includegraphics[width=0.35\textwidth]{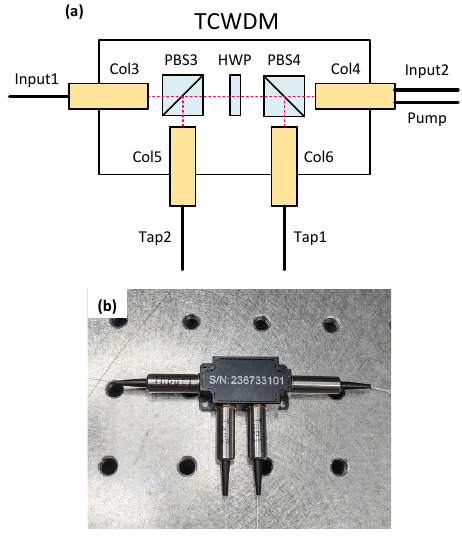}
	\caption{The principle diagram (a) and a picture (b) of the custom-designed TCWDM. Col3, Col5 and Col6, collimators; Col4, WDM collimator, PBS3 and PBS4, polarization beam splitters; HWP, half-wave plate.}
	\label{fig2}
\end{figure}

We designed a tap coupler hybrid with WDM (TCWDM, AFR) with the fast axis blocked. The principle diagram and a picture of the TCWDM are shown in Fig.~\ref{fig2}. The tap ratio of the TCWDM is 10\% and the pump light in the WDM collimator (Col4) is reflected to the Input2 port. The package length of the TCWDM is only 64~mm, which is shorter than that of the fiber device made by a fiber coupler and a WDM. In Fig.~\ref{fig2}(a), the polarization of the slow-axis light from Col3 is parallel to the $p$ state of the PBS3, and that from Col4 is parallel to the $p$ state of the PBS4. Thus, the angle of the half-wave plate (HWP) determines the tap ratio of the TCWDM.

The following outlines our in-situ alignment method based on the custom-designed TCWDM.
\begin{itemize}
	\item[$\mathbf{a.}$] Splice the fibers and set up the optical path as shown in Fig.~\ref{fig1}, and remove FR, QWP, and PBS2.
	\item[$\mathbf{b.}$] Introduce the external linearly polarized light from Tap1 and Tap2, respectively, and rotate Col1 and Col2 to align their slow axes in the desired direction based on the power of light passing through or reflecting from PBS1.
	\item[$\mathbf{c.}$] Adjust the angle and height of Col1 to make the output light horizontal and aligned with the center of PBS1 and fix Col1. Introduce light from Tap1, and obtain ideal coupling efficiency between Col1 and the mirror by adjusting the mirror according to the output power of Tap1.
	\item[$\mathbf{d.}$] Introduce light from Tap2, and obtain maximum coupling efficiency between Col2 and the mirror by adjusting the Col2 based on the output power of Tap2.
	\item[$\mathbf{e.}$] Place the FR at the position shown in Fig.~\ref{fig1}, provide the pump to the gain fiber, and obtain ideal coupling efficiency between Col1 and Col2 by adjusting the Col2 based on the output power of Tap1.
\end{itemize}

In the final step, the coupling between Col1 and Col2 is not realized by introducing the external light from Tap1 or Tap2. The reason is as follows. When the external light is introduced from Tap1, 90\% of the input light is emitted from Tap2 and 10\% remains in the loop. After experiencing the splitting by a fiber coupler, the absorption by the gain fiber, and the collimator coupling loss, the remaining light will emit from Tap2, but with a power much lower than 1\% of the input light. This low power light is used to identify the coupling between the two collimator. Therefore, it requires sufficient stability of the light source power. Furthermore, interference between the two components of the output light from Tap2 can lead to significant power fluctuations, which makes it unsuitable for identifying the efficiency of the coupling.

\begin{figure}[!t]
	\centering
	\includegraphics[width=0.43\textwidth]{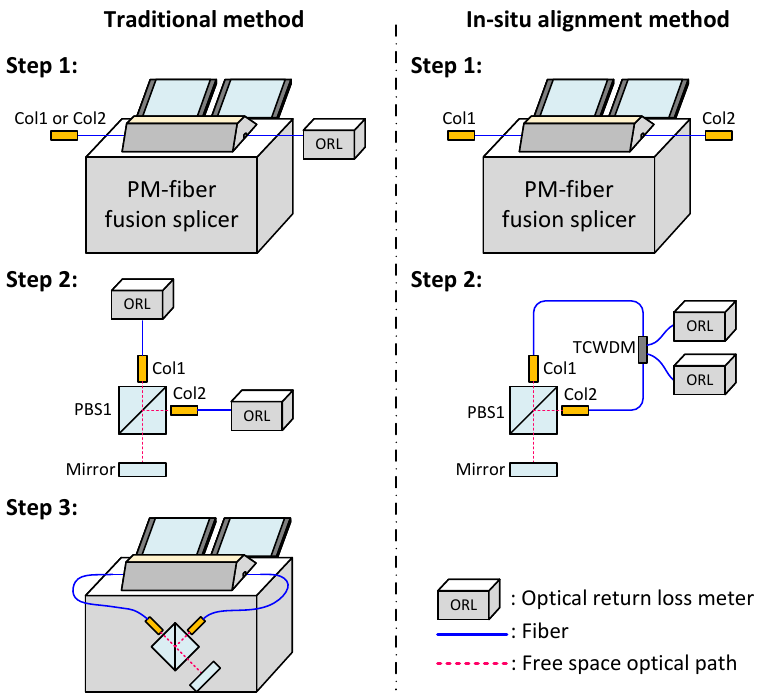}
	\caption{Comparison between traditional method and in-situ alignment method.}
	\label{fig3}
\end{figure}

Our in-situ alignment method has certain advantages compared to the traditional method. In the traditional method, as shown in Fig.~\ref{fig3}, the all-PM fiber laser mode-locked by a NALM is typically built in three steps. The collimators are coupled when the loop is open, involving disconnecting the tail fibers and splicing fiber for several times. 
At least around 65 cm tail fiber must be reserved for splicing due to the following facts: (1) The two collimators in step 3 of Fig.~\ref{fig3} are close together and even mounted on the optical table through collimator fixtures and six-axis alignment stages. (2) The size of the PM-fiber fusion splicer (such as FSM-100P, Fujikura) is large. (3) The fiber fixtures have large sizes and need to be rotated.
The tail fibers of fiber devices in high-repetition-rate lasers are only enough to be spliced a few times. If the collimator coupling efficiency decreases due to environmental disturbances affecting the fabricated laser, all fiber devices may need to be scrapped when re-coupling. 
The in-situ alignment method can avoid the shortcomings of the traditional method, as the fibers are linearly placed in the fusion splicer, as shown in Fig.~\ref{fig1} and Fig.~\ref{fig3}. The fiber fusion splicing process is straightforward, similar to that of linear cavity mode-locked fiber lasers~\cite{lv2018,liu2023}. Our in-situ alignment method relies on the laser itself and the reserved tail fiber can be shortened by nearly half (about 20--30~cm). Note that the alignment method that uses an external reference collimator to assist in coupling collimators is not discussed in the paper, as the coupling between collimators and reflectors also requires a longer cavity length or complex alignment system.

\section{Results and Discussion}
\begin{figure}[!t]
	\centering
	\includegraphics[width=0.45\textwidth]{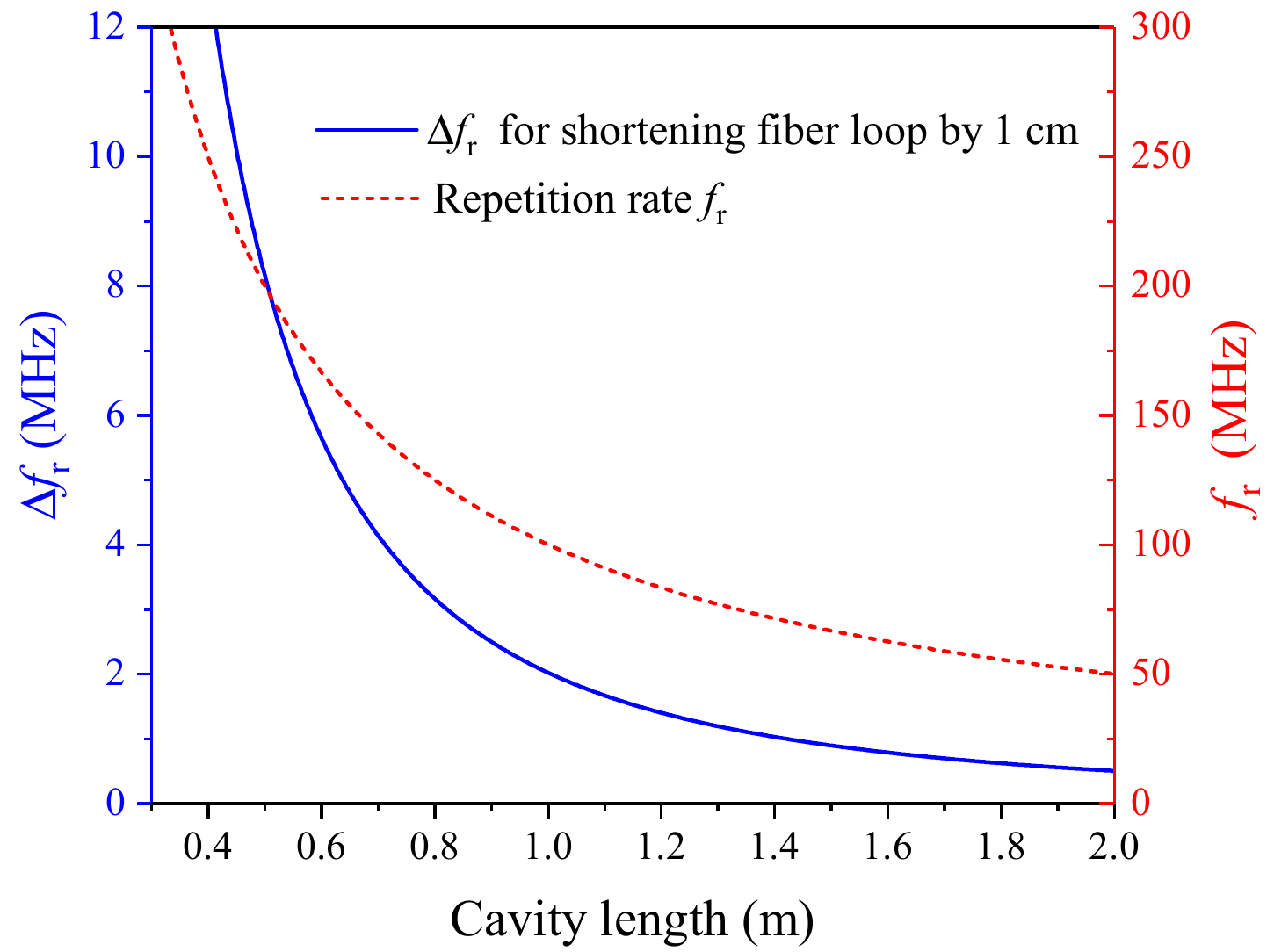}
	\caption{Repetition rate changes when the fiber loop is shortened by 1 cm at different cavity lengths.}
	\label{fig4}
\end{figure}

The repetition rate change caused by shortening fiber loop by 1~cm is
\begin{equation}
	\varDelta f_r=\frac{c}{n}\left( \frac{1}{2L-0.01}-\frac{1}{2L} \right) .
\end{equation}
Here, the 1~cm length refers to the fiber segment that needs to be cut during fusion splicing, $c$ is the speed of light, $n$ is the refractive index of the fiber, $L$ is the cavity length. The contribution of the length of free space is ignored. The repetition rate $f_r$ and the $\varDelta f_r$ are shown in Fig.~\ref{fig4}. It can be seen that as $f_r$ increases from 100~MHz to 200~MHz, $\varDelta f_r$ caused by shortening the fiber loop by 1~cm increases by 1.5~MHz. When $f_r$ increases from 200~MHz to 300~MHz, $\varDelta f_r$ increases by 6~MHz. Therefore, the fiber shortened by about 20--30~cm in the in-situ alignment method can greatly improve $f_r$. The $\varDelta f_r$ in free space is slightly different from that of the fiber loop, and the cavity-length relationship is
\begin{equation}
	2n'\varDelta l'=n\varDelta l.
\end{equation}
Here, $\varDelta l'$ and $\varDelta l$ represent the changes in cavity length in free space and fiber loop, respectively; and $n'$ is the refractive index of air. Thus, 1~cm change in the cavity length of free space is equivalent to 1.3~cm change in fiber loop, which indicates that compactness is also important in improving the repetition rate.

\begin{figure}[!t]
	\centering
	\includegraphics[width=0.45\textwidth]{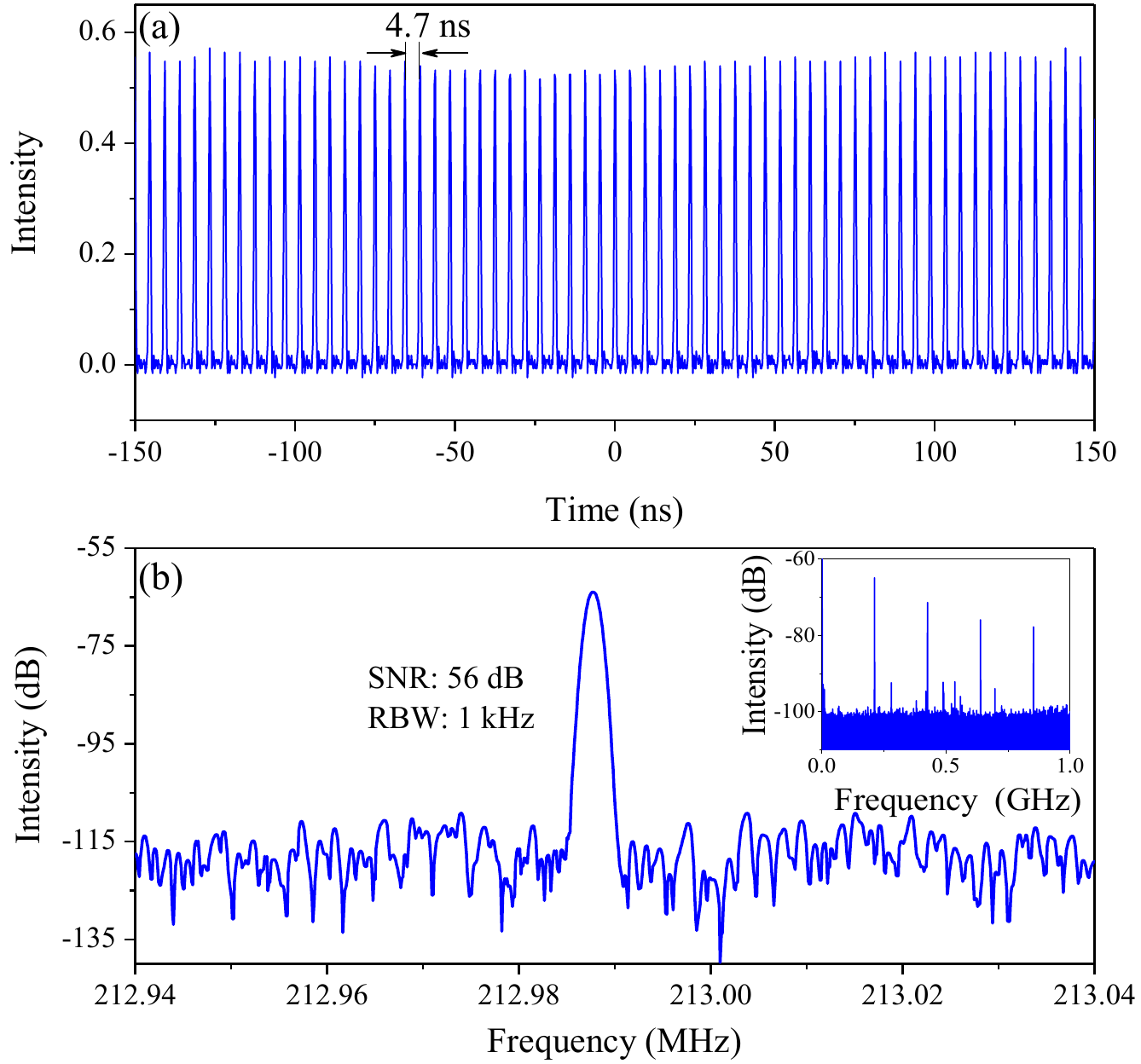}
	\caption{(a) Oscilloscope trace of the pulses emitted from Tap2. (b) RF spectra of the pulses emitted from Tap2. Inset: the broad-span RF spectrum.}
	\label{fig5}
\end{figure}

The mode-locked laser operates at a repetition rate of 213~MHz with a signal-to-noise ratio (SNR) of 56~dB at a resolution bandwidth (RBW) of 1~kHz, as shown in Fig.~\ref{fig5}. The pulses emitted from Tap2 are detected by an InGaAs biased detector (DET08CFC, Thorlabs), and the pulse interval displayed on the oscilloscope (DSO-X 3054A, Agilent) is 4.7~ns, corresponding to the repetition rate on the spectrum analyzer (N9020A, Keysight). At 1300~mW pump power, the output powers of the four ports are presented in Table~\ref{tab1}. As a result of experiencing gain fiber amplification earlier, the output power of Tap2 is larger than that of Tap1. When varying the pump power while maintaining mode-locking, the output power of the Col2 port remains the highest, and the proportional relationship of the output powers of the four output ports changes slightly.

\begin{table}[b]
	\caption{\label{tab1}%
		Output powers of four output ports of the laser at 1300~mW (mode-locking) and 1240~mW (non-mode-locking) pump powers.
	}
	\begin{ruledtabular}
		\begin{tabular}{lcc}
			\textrm{Output port}&\textrm{Mode-locking}&\textrm{Non-mode-locking}\\
			\colrule
			Col2 port& 41.3 mW & 33.5 mW \\
			PBS2 port& 22.8 mW & 66.1 mW \\
			Tap1 & 2.6 mW & 1.7 mW \\
			Tap2 & 14.1 mW & 12.3 mW \\
		\end{tabular}
	\end{ruledtabular}
\end{table}

\begin{figure}[!t]
	\centering
	\includegraphics[width=0.43\textwidth]{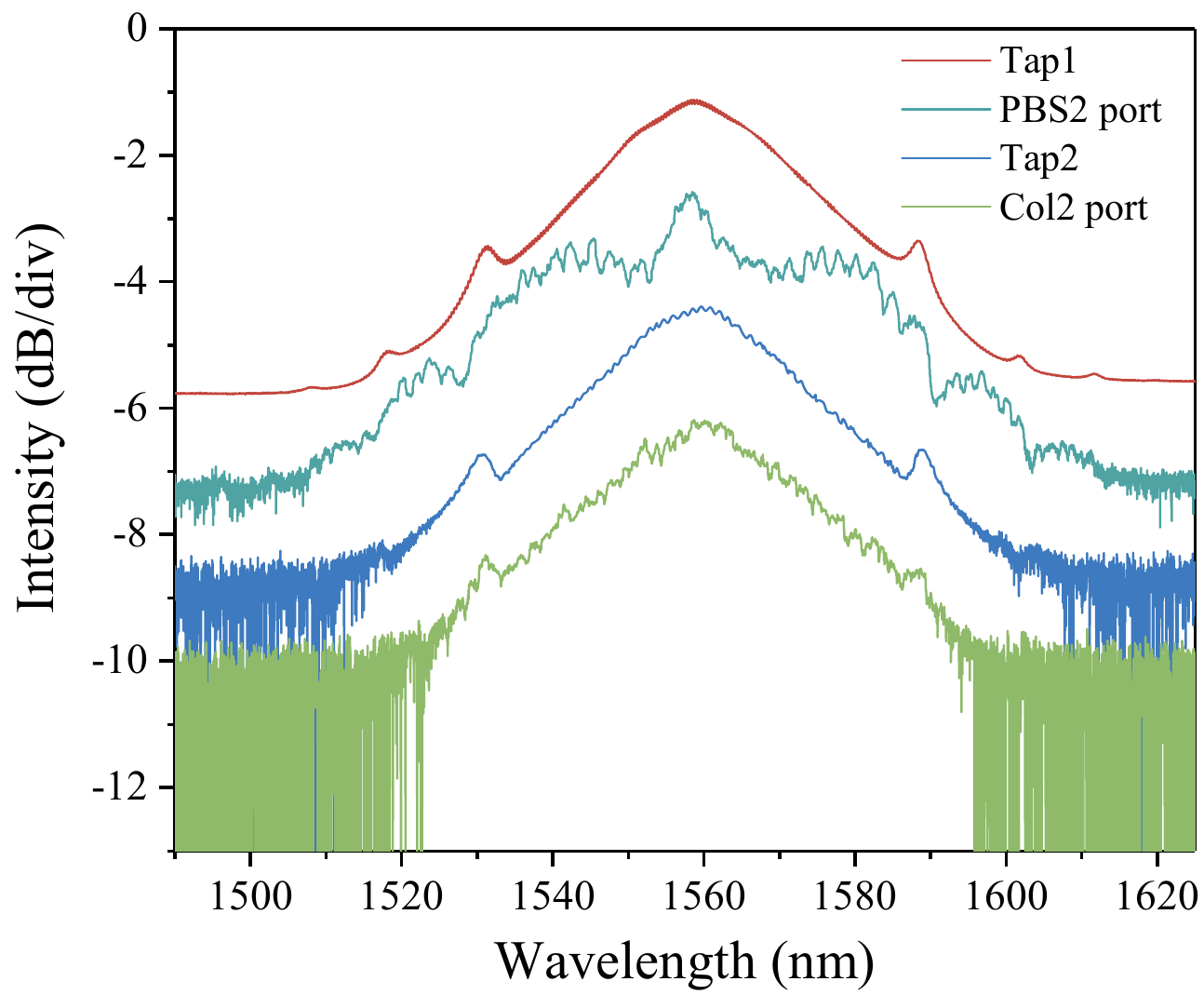}
	\caption{Optical spectra of the four output ports of the mode-locked laser. The spectral bases are 1.35 dB apart.}
	\label{fig6}
\end{figure}

We employ an optical spectrum analyzer (MS9710C, Anritsu) to measure the spectra of the four output ports of the mode-locked laser, as shown in Fig.~\ref{fig6}. It is found that the pulses emitted from Tap1 have the smoothest spectrum, featuring three pairs of symmetrical Kelly sidebands~\cite{kelly1992}. The pulses emitted from Tap2 have a slightly less smooth spectrum with noticeable background noise, which is attributed to negative-GVD erbium-doped fiber amplification (not chirped-pulse amplification). The spectrum of the Col2 port exhibits reduced smoothness, and its background noises on both sides are closer to the center. This is caused by the combination of pulses in the fast and slow axes and the continuous-wave light generated by amplified spontaneous emission. The spectrum of the PBS2 port is roughly similar to that of the Tap2 port, and the major difference lies in the large peak at the center. This is because the energy of two interfering beams of light at this peak is significantly different. If a QWP is placed between PBS2 and the mirror, a corresponding dip will appear in the spectrum of the light emitted from the other side of PBS2~\cite{laszczych2022}.

Except for Kelly sidebands, the small spikes in the spectrum of the PBS2 port result from the energy difference at corresponding wavelengths when the two beams emitted from two collimators interfere. The spectral spikes are caused by erbium-doped fiber amplification; this significantly contributes to this energy difference, rather than by the widely believed large nonlinear effects. This is because the spectrum before interference, i.e., the spectrum of the Col2 port, differs greatly from that of the PBS2 port, as there are almost no nonlinear effects in the free-space optical path. From the spectra in Fig.~\ref{fig6}, we can obtain the spectral evolution relationships as follows: (1) The light interference on PBS2 leaves interference traces in the spectrum of the PBS2 port. (2) The traces are cleaned up after reflection and passing through a nonreciprocal phase shifter and two PBSs, resulting in the cleanest spectrum light after splitting into the fiber loop. This light here is suitable for output, and this output can affect the saturable absorption effects of NALM. (3) The spectrum of the light entering the passive fiber remains clean, but after erbium-doped fiber amplification, the spectrum will exhibit spikes and background noise. Then the light returns to the PBS1 port with a poor-quality spectrum.

\begin{figure}[!t]
	\centering
	\includegraphics[width=0.43\textwidth]{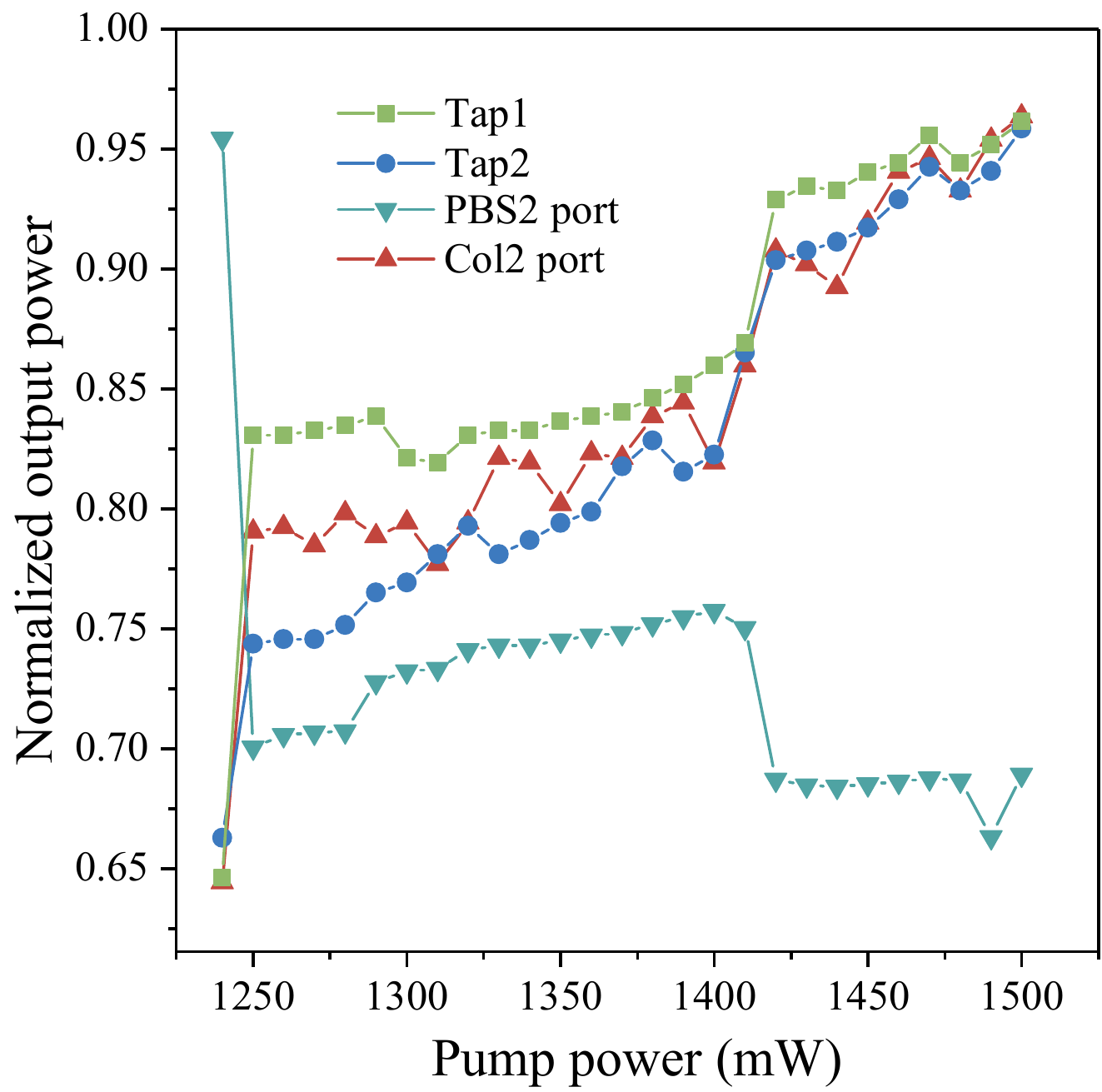}
	\caption{Changes in normalized output power of four output ports of the mode-locked laser as pump power gradually decreases.}
	\label{fig7}
\end{figure}

The pulses from the four output ports of the mode-locked laser are related not only in the spectrum but also in output power. By gradually reducing the pump power in 10~mW intervals after achieving single-pulse mode-locking at the highest available 1500~mW pump power, the output power evolutions are obtained as shown in Fig.~\ref{fig7}. We use the normalization method to bring the magnitudes of the output powers closer. Previously, some researchers also provided power variation diagrams, but they were either for a single output port or for multiple output ports without normalization, making it difficult to observe the relationships of the output powers. In Fig.~\ref{fig7}, the overall variation trend of the output power of the PBS2 port is opposite to the other three output ports. This indicates that when the pump power is reduced, the output loss of NALM will actually increase. In terms of the smoothness of the curve changes, the Tap1 and PBS2 curves are the smoothest, the Tap2 curve is the next, while the Col2 curve is the worst. This is because the light transmitted from PBS2 is reflected to the fiber loop and then is emitted from Tap1, while the lights from the Col2 port and the Tap2 port are amplified by the gain fiber. When the pump power decreases to around 1430~mW, the output power of the PBS2 port suddenly increases while the output power of the other three output ports suddenly decreases. This is because, similar to the multi-pulse regime, there are multistability states in the single-pulse regime, and the sudden change of the output power is caused by a transition between different stability states~\cite{komarov2005,tang2005}. Additionally, when the pump power decreases to 1240~mW, mode-locking cannot be maintained, and the output powers of the four output ports are shown in Table~\ref{tab1}. It can be seen that the output power of the PBS2 port suddenly increases significantly while the output powers of the other three output ports decrease with a similar scale. This difference is caused by the cavity structure of the Figure-9 laser itself, which results in a large variation in output power of the PBS2 port with and without saturable absorption effects, and this can also be used to judge the mode-locking of the laser.

\section{Conclusion}
We propose an in-situ alignment method to realize an all-PM erbium-doped fiber laser mode-locked by a NALM. Although a relatively low-gain fiber and two taps of the fiber loop are employed, a repetition rate of 213~MHz is still achieved. Compared to the traditional method, the in-situ alignment method not only separates the fiber fusion-splicing from coupling collimators but also avoids the spatial position limitations of the fusion splicer and well-coupled collimators. Moreover, when the coupling efficiency of the collimator decreases, it can be re-adjusted in situ. If non-standard optical devices and a larger number of pump semiconductor lasers are used, a higher repetition rate can be obtained. Through measuring and comparing the spectra of the four output ports of the mode-locked laser, we observe the spectral interference phenomena and provide the evolution relationship and spectral quality ranking among different output ports. It is widely believed that spectral distortion is caused by large negative net cavity dispersion (large nonlinear effects), however, our findings indicate that light interference is also a significant factor. In addition, by analyzing the relationships between pump powers and output powers of the four output ports, we find that a single-pulse regime contains multiple stability states.

\smallskip{}

\begin{acknowledgments}
This work was supported in part by National Natural Science Foundation of China under Grants 62105368, 62275268 and 62375284, and in part by The Science and Technology Innovation Program of Hunan Province under Grant 2023RC3010.
\end{acknowledgments}

\end{document}